\documentclass[seceq]{ptptex}

\usepackage{wrapft}
\usepackage{amsfonts}
\usepackage{latexsym,cite,amssymb}
\usepackage{times}
\usepackage{graphicx}
\usepackage{color}

\makeatother

\newcommand{\bea}{\begin{eqnarray}}
\newcommand{\eea}{\end{eqnarray}}
\newcommand{\be}{\begin{equation}}
\newcommand{\ee}{\end{equation}}
\newcommand{\bem}{\begin{pmatrix}}
\newcommand{\eem}{\end{pmatrix}}
\newcommand{\bl}{\begin{align}}
\newcommand{\el}{\end{align}}

\title{%        %You can use \\ for explicit line-break.
Analytical calculation on critical magnetic field in holographic  superconductors with backreaction
}

\author{%       %Use \scshape  for the family name.
Xian-Hui \textsc{Ge}$^{1,2,}$\footnote{E-mail: gexh@shu.edu.cn}
and Hong-Qiang \textsc{Leng}$^{1,}$\footnote{E-mail: lenghq88@shu.edu.cn}
}

\inst{%         %Affiliation, neglected when [addenda] or [errata].
$^1$Department of Physics, Shanghai University, Shanghai 200444, China\\
$^2$Kavli Institute for Theoretical Physics China, CAS, Beijing 100190, China
}

\abst{%         %This abstract is neglected when [addenda] or [errata].
We investigate the effect of spacetime backreaction on the upper critical magnetic field for $s$-wave holographic superconductors by using the matching method.
The backreaction of the constant external magnetic field and the electric field to the background geometry leads to a dyonic black hole solution.
The magnetic fields as well as the electric fields acting as gravitational sources tend to depress the critical temperature of the superconductor.
We derive the analytical expression for the upper critical magnetic field up to $\mathcal{O}(\kappa^2)$ order and find that backreaction makes the upper critical magnetic field stronger.  The result is consistent with the previous numerical and analytical results.
}

\begin{document}
\maketitle

\section{Introduction}
The gauge/gravity duality  \cite{ads/cft,gkp,w} as  the most fruitful idea stemming from string theory, has been proved to be a powerful tool for studying the strongly coupled systems in field theory. By using a dual classical gravity description, we can effectively calculate correlation functions   in a strongly interacting field theory.
 Recently, a superconducting phase was established with the help of black hole physics in higher dimensional spacetime\cite{gub1,gub2,3h,horowitz}.

Counting on the numerical calculations, the critical temperature was calculated with and without the backreaction for various conditions \cite{horowitz2,gre,wang1,wang2,wang3,wang4,siani,wang5,cai,ling,kuang,hartman,barc,kanno,jing,chen,amm1}.
The behavior of holographic superconductors in the presence of  an external magnetic field has been widely studied in the probe limit\cite{aj,wk,amm,john,mon,ns,silva,gw,wu,queen,mann}.
  The analytical calculation is useful for gaining insight into the strong interacting system. If the problem can be solved analytically, however vaguely, one can usually gain some insight.
As an analytical approach for deriving the upper critical magnetic field, an expression  was found in the probe limit by  extending the matching method first proposed in \cite{gre} to the magnetic case \cite{gw},
which is shown to be consistent with the Ginzburg-Landau theory.

 Most of previous studies on the holographic superconductors  focus on the probe limit neglecting  the backreaction of matter field on the spacetime.
 The probe limit corresponds to the case the electric charge $q\rightarrow \infty$ or the Newton constant  approaches zero.
 The backreaction of the spacetime becomes important in the case away from the probe limit. At a lower temperature, the black hole becomes hairy and the phase diagram might be modified. Recently, an analytical calculation on the critical temperature of the Gauss-Bonnet holographic superconductors with backreaction was presented in \cite{kanno} and confirmed the numerical results that
  backreaction makes condensation harder\cite{hartman,barc,wang4,siani} .

  Considering the above facts, it would be of great interest to explore the behavior of the upper critical magnetic field for holographic superconductors in the presence of the backreaction.
  In this work, we will first consider the effect of the spacetime backreaction to $s$-wave holographic superconductors without the magnetic field. Different from the probe limit case, the backreaction of spacetime actually leads to a charged black hole solution in AdS space at the leading order. We will compute the critical temperature analytically by using this charged black hole metric through the matching method. Secondly we will study the properties of holographic superconductors in the presence of external magnetic field.   When we turn on the external magnetic field, the resulting background geometry becomes the dyonic black hole solution in AdS space to the zeroth order. The analytical investigation on the effect of the spacetime backreaction to the upper critical magnetic field  has not been  carried out so far. Therefore, the contents in this paper will be greatly different from the probe limit case as we consider the spacetime backreaction. Note that in both cases, the small backreaction approximation shall be used to obtain an analytical result. In order to compare the analytic study with numerical results, we will also carry on numerical computation.

The organization of the paper is as follows: We first consider the effects of backreaction on $2+1$-dimensional $s$-wave holographic superconductors in section 2. Without the magnetic field, the critical temperature  with backreaction  will be derived first. Then we continue the calculation to the strong external magnetic field case and find an analytical expression for the backreaction on the upper critical magnetic field.
{From the Einstein equation, we know that the presence of charge and magnetism in $4$-dimensional spacetime yields a dyonic black hole solution. The critical temperature may influenced by the backreaction of the magnetism.  We will compare the analytic and numerical results.
 We present the conclusion will be presented in the last section.}

\section{$(2+1)$-dimensional $s$-wave holographic superconductors}

{In this section, we first investigate the backreaction of electric field on superconductivity and derive the phase transition temperature $T_c$ in this case. After that, we turn to the backreaction of the external magnetic field and calculate the critical magnetic field.
}

\subsection{Critical temperature with backreaction in the absence of magnetic fields}

We begin with  a charged, complex scalar field into the 4-dimensional
Einstein-Maxwell action with a negative cosmological constant
\begin{equation}\label{action}
S=\frac{1}{16\pi G_{4}}\int d^4
x\sqrt{-g}\bigg\{R-2\Lambda-\frac{1}{4}F_{\mu\nu}F^{\mu\nu}
-|\partial_{\mu}\psi-iqA_{\mu}\psi|^2-m^2|\psi|^2\bigg\},
\end{equation}
where $G_{4}$ is the 4-dimensional Newton constant, the cosmological
constant $\Lambda=-3/l^2$ and
$F_{\mu\nu}=\partial_{\mu}A_{\nu}-\partial_{\nu}A_{\mu}$.
The hairy black hole solution is assumed to take the following metric ansatz
\be
ds^2=-f(r)e^{-\chi(r)}dt^2+\frac{dr^2}{f(r)}+\frac{r^2}{l^2}(dx^2+dy^2),
\ee
together with
\be
A_{\mu}=(\phi(r),0,0,0),~~~\psi=\psi(r).
\ee
The Hawking temperature, which will be interpreted as the temperature of the holographic superconductors, is given by
\be
T=\frac{1}{4\pi}f'(r)e^{-\chi(r)/2}\bigg|_{r=r_{+}},
\ee
where a prime denotes derivative with respect to $r$ and $r_{+}$ is the black hole horizon defined by $f(r_{+})=0$. $\psi(r)$ can be taken to be real by using the $U(1)$ transformation. The gauge and scalar equations become
\bea
&&\phi''+(\frac{\chi'}{2}+\frac{2}{r})\phi'-\frac{2q^2\psi^2}{f}\phi=0,\\
&&\psi''+(\frac{f'}{f}-\frac{\chi'}{2}+\frac{2}{r})\psi'+\bigg(\frac{q^2\phi^2 e^{\chi}}{f^2}-\frac{m^2}{f}\bigg)\psi=0.
\eea
The $tt$ and $rr$ components of the background Einstein equations yield
\bea
&&f'+\frac{f}{r}-\frac{3r}{l^2}+\kappa^2 r\bigg[\frac{e^{\chi}}{2}\phi'^2+m^2\psi^2+f\bigg(\psi'^2+\frac{q^2\phi^2\psi^2e^{\chi}}{f^2}\bigg)\bigg]=0,\\
&&\chi'+2\kappa^2r\bigg(\psi'^2+\frac{q^2\phi^2\psi^2e^{\chi}}{f^2}\bigg)=0.
\eea
When the Hawking temperature is above a critical temperature, $T>T_c$ the solution is the well-known AdS-Reissner-Nordstr$\ddot{o}$m (RNAdS) black holes
\be
f=\frac{r^2}{l^2}-\frac{1}{r}\bigg(\frac{r^3_{+}}{l^2}+\frac{\kappa^2\rho^2}{2 r_{+}}\bigg)+\frac{\kappa^2\rho^2}{2r^2}, ~~~\chi=\psi=0,~~~\phi=\rho\bigg(\frac{1}{r_{+}}-\frac{1}{r}\bigg),
\ee
where $\kappa^2=8\pi G_4$. {At the critical temperature $T=T_c$, the coupling of the scalar to gauge field induces an effective negative
mass term for the scalar field,the RNAdS solution thus becomes unstable against perturbation of the scalar field.}
At the asymptotic AdS boundary ($r\rightarrow \infty$), the scalar and the Maxwell fields behave as
\be
\psi=\frac{<\mathcal{O}_{\Delta_{-}}>}{r^{\Delta_{-}}}+\frac{<\mathcal{O}_{\Delta_{+}}>}{r^{\Delta_{+}}}, ~~~~~\phi=\mu-\frac{\rho}{r}+...
\ee
where $\mu$ and $\rho$ are interpreted as the chemical potential and charge density of the dual field theory on the boundary. According to the gauge/gravity duality,
 $<\mathcal{O}_{\Delta_{\pm}}>$ represents the expectation value of the operator $\mathcal{O}_{\Delta_{\pm}}$ dual to the charged scalar field $\psi$.
The exponent $\Delta_{\pm}$ is determined by the mass as $\Delta_{\pm}=\frac{3}{2}\pm \frac{1}{2}\sqrt{9+4m^2}$. Note that for $\psi$ both of the falloffs are normalizable and we choose the boundary condition that
either $<\mathcal{O}_{\Delta_{-}}>$ or $<\mathcal{O}_{\Delta_{+}}>$ is vanishing. We will impose that $\rho$ is fixed and take $<\mathcal{O}_{\Delta_{-}}>=0$ as in \cite{horowitz}.   Moreover, we will consider the values of $m^2$ which must satisfy the Breitenlohner-Freedman (BF) bound $m_{BF}^2  \leq  m^2  <  m_{BF}^2 + 1$ with $m_{BF}^2  = - (d-1) / 4$ \cite{BF} for the dimensionality of the spacetime $d=4$ in the following analysis.

After introducing the new coordinate $z=\frac{r_{+}}{r}$, the equations of motion become
\bea
&&-f'+\frac{f}{z}-\frac{3r^2_{+}}{l^2z^3}+\kappa^2\frac{r^2_{+}}{z^3}\bigg[\frac{z^4e^{\chi}}{2r^2_{+}}\phi'^2+m^2\psi^2
+f(\frac{z^4}{r^2_{+}}\psi'^2+\frac{q^2\phi^2\psi^2e^{\chi}}{f^2})\bigg]=0,\label{m1}\\
&&-\chi'+2\kappa^2\frac{r^2_{+}}{z^3}\bigg[\frac{z^4}{r^2_{+}}\psi'^2+\frac{q^2\phi^2\psi^2e^{\chi}}{f^2})\bigg]=0,\label{m2}\\
&&\phi''+\frac{1}{2}\chi'\phi'-\frac{2q^2\psi^2r^2_{+}}{z^4f}\phi=0,\label{at}\\
&&\psi''-\bigg(\chi'-\frac{f'}{f}\bigg)\psi'+\frac{r^2_{+}}{z^4}\bigg(\frac{q^2\phi^2 e^{\chi}}{f^2}-\frac{m^2}{f}\bigg)\psi=0, \label{p1}
\eea
where the prime $'$ denotes a derivative with respect to $z$.  One may find that the transformation $\phi\rightarrow \phi/q$ and $\psi\rightarrow \psi/q$ does not change the form of the Maxwell and the scalar equations, but the gravitational coupling of the Einstein equation changes $\kappa^2\rightarrow \kappa^2/q^2$. The probe limit studied in \cite{horowitz} corresponds to the limit $q\rightarrow \infty$ in which the matter sources drop out of the Einstein equations. The hairy black hole solution requires to go beyond the probe limit. In \cite{horowitz},  it was suggested to  take  finite $q$ by setting $2\kappa^2=1$. Recently, the author in \cite{kanno}  proposed to keep
$2\kappa^2$ finite with setting $q=1$ instead. We will take the latter choice.

{In the neighborhood of the critical temperature $T_c$,we can choose the order parameter as an expansion parameter because it is small valued}
\be
\epsilon\equiv <\mathcal{O}_{\Delta_{+}}>.
\ee
{We find that given the structure of our equations of motion, only the even orders of $\epsilon$ in the gauge field and gravitational field,
and odd orders of $\epsilon$ in the scalar field appear here. That is to say, we  can expand the scalar field $\psi$, the gauge field as a series in $\epsilon$ as}
\bea
&&\phi=\phi_0+\epsilon^2 \phi_2+\epsilon^4 \phi_4+...\\
&&\psi=\epsilon \psi_1+\epsilon^3 \psi_3+\epsilon^5 \psi_5+...
\eea
{Let us expand the background geometry line elements $f(z)$ and $\chi(z)$  around the AdS-Reissner-Nordstr$\ddot{o}$m solution}
\bea
&&f=f_0+\epsilon^2 f_2+\epsilon^4 f_4+...\\
&&\chi=\epsilon^2 \chi_2+\epsilon^4 \chi_4+...
\eea
{The chemical potential $\mu$ should also expanded as}
\be
\mu=\mu_0+\epsilon^2 \delta\mu_2,
\ee
where $\delta\mu_2$ is positive. {Therefore, near the phase transition, the order parameter as a function of the chemical potential, has the form}
\be
\epsilon=\bigg(\frac{\mu-\mu_0}{\delta\mu_2}\bigg)^{1/2}.
\ee
 {It is clear that  that when $\mu$ approaches $\mu_0$, the order parameter $\epsilon$ approaches zero. The phase transition  occurs at the critical value $\mu_c=\mu_0$. Note that the critical exponent $1/2$ is the universal result from the Ginzburg-Landau mean field theory. The equation of motion for $\phi$ is solved
at zeroth order by $\phi_0=\mu_0(1-z)$ and this gives a relation $\rho=\mu_0r_{+}$. So, to zeroth order the equation for $f$ is solved as}
\be
f_0(z)=\frac{r^2_{+}}{z^2l^2}\bigg(1-z\bigg)\bigg(1+z+z^2-\frac{\kappa^2l^2\mu^2_0}{2r^2_{+}}z^3\bigg).
\ee
Now the horizon locates at $z=1$. We will see that the critical temperature with spacetime backreaction can be determined by solving the equation of motion for $\psi$ to the first order.

At first order, we need solve the equation for $\psi_1$ by the matching method. The boundary condition and regularity at the horizon requires
\be
\psi'(1)=\frac{r^2_{+}m^2}{f'_0(1)}\psi_1(1).\label{psi1}
\ee
In the asymptotic AdS region, $\psi_1$ behaves like
\be
\psi_1=C_{+}z^{\Delta_{+}},\label{as}
\ee
Now let us expand $\psi_1$ in a Taylor series near the horizon
\be
\psi_1=\psi_1(1)-\psi'_1(1)(1-z)+\frac{1}{2}\psi''_1(1)(1-z)^2+...\label{pp}
\ee
From (\ref{p1}), we obtain the second derivative of $\psi_1$ at the horizon
\be
\psi''_1(1)=-\frac{1}{2}\bigg(4+\frac{f''_0(1)}{f'_0(1)}-\frac{m^2r^2_{+}}{f'_0(1)}\bigg)\psi'_{1}(1)-\frac{r^2_{+}\phi'_0(1)^2}{2 f'^2_0(1)}\psi_1(1).\label{psi2}
\ee
Using (\ref{psi1}) and (\ref{psi2}), we find the approximate solution near the horizon
\be
\psi_1(z)=\psi_1(1)-\frac{r^2_{+}m^2}{f'_0(1)}\psi_1(1)(1-z)+\bigg[-\frac{r^2_{+}m^2}{4f'_0(1)}\bigg(4+\frac{f''_0(1)}{f'_0(1)}-\frac{r^2_{+}m^2}{f'_0(1)}\bigg)
-\frac{r^2_{+}}{4}\frac{\phi'_1(1)^2}{f'_0(1)^2}\bigg]\psi_1(1)(1-z)^2+...
\ee
In order to determine $\psi_1(1)$ and $C_{+}$, we match the solution (\ref{as}) and (\ref{pp}) smoothly at $z_m$. We find that
\bea
z^{\Delta_{+}}_m C_{+}&=&\psi_1(1)-\frac{r^2_{+}m^2}{f'_0(1)}\psi_1(1)(1-z_m)+\bigg[-\frac{r^2_{+}m^2}{4f'_0(1)}\bigg(4+\frac{f''_0(1)}{f'_0(1)}-\frac{r^2_{+}m^2}{f'_0(1)}\bigg)
\nonumber\\&&-\frac{r^2_{+}}{4}\frac{\phi'_1(1)^2}{f'_0(1)^2}\bigg]\psi_1(1)(1-z_m)^2,\\
\Delta_{+}z^{\Delta_{+}-1}_mC_{+}&=&\frac{r^2_{+}m^2}{f'_0(1)}\psi_1(1)-2\bigg[-\frac{r^2_{+}m^2}{4f'_0(1)}\bigg(4+\frac{f''_0(1)}{f'_0(1)}-\frac{r^2_{+}m^2}{f'_0(1)}\bigg)
\nonumber\\&&-\frac{r^2_{+}}{4}\frac{\phi'_1(1)^2}{f'_0(1)^2}\bigg]\psi_1(1)(1-z_m).\label{d}
\eea
Solving the above equation, we obtain the expression for $C_{+}$ in terms of $\psi_1(1)$
\be
C_{+}=\frac{2z_m}{2z_m+(1-z_m)\Delta_{+}}z^{-\Delta_{+}}_m \bigg(1-\frac{1-z_m}{2}\frac{r^2_{+}m^2}{f'_0(1)}\bigg)\psi_1(1).
\ee
Substituting the above equation back into (\ref{d}), we get a non-trivial relation provided $\psi_1(1)\neq 0$,
\bea
&&\frac{2\Delta_{+}}{2z_m+(1-z_m)\Delta_{+}}-
\bigg[\frac{(1-z_m)\Delta_{+}}{2z_m+(1-z_m)\Delta_{+}}+(3-2z_m)\bigg]\frac{r^2_{+}m^2}{f'_0(1)}\nonumber\\&-&\frac{(1-z_m)r^2_{+}m^2}{2}\frac{f''_0(1)}{f'_0(1)^2}
+\frac{1-z_m}{2}\frac{r^4_{+}m^4}{f'_0(1)^2}-\frac{(1-z_m)r^2_{+}}{2}\frac{\phi'_0(1)^2}{f'_0(1)^2}=0.\label{re1}
\eea
Note that $f'_0(1)=-\frac{r^2_{+}}{l^2}\bigg(3-\frac{\kappa^2l^2\mu^2_0}{2r^2_{+}}\bigg)$, $f''_0(1)=\frac{r^2_{+}}{l^2}\bigg(6+\frac{\kappa^2l^2\mu^2_0}{r^2_{+}}\bigg)$ and $\phi'_0(1)=-\mu_0$. Plugging these relations back into
(\ref{re1}), we obtain an equation for $\mu_0$
\bea\label{main}
&&\frac{\kappa^4l^4}{2r^4_{+}}\frac{\Delta_{+}}{2z_m+(1-z_m)\Delta_{+}}\mu^4_0-\frac{l^4(1-z_m)}{2r^2_{+}}\bigg\{1+2\kappa^2\frac{r^2_{+}}{l^4(1-z_m)}
\bigg[\frac{m^2l^4(1-z_m)\Delta_{+}}{2r^2_{+}(2z_m+(1-z_m)\Delta_{+})}\nonumber\\&&+\frac{6\Delta_{+}l^2}{r^2_{+}(2z_m+(1-z_m)\Delta_{+})}
-\frac{m^2l^4}{r^2_{+}}\bigg(\frac{3}{2}z_m-2\bigg)\bigg]\bigg\}\mu^2_0\nonumber\\&&+3m^2l^2(2-z_m)+\frac{m^4l^4}{2}(1-z_m)-\frac{18+3m^2l^2(1-z_m)}{z_m(\Delta_{+}-2)-\Delta_{+}}\Delta_{+}=0.
\eea
The main idea of \cite{kanno} is to work in the small backreaction approximation $\kappa^2\ll 1$ together with the matching method so that all the functions can be expanded by $\kappa^2$ and the $\kappa^4$ term in the above equation
can be neglected. In this sense,  $\mu_0$ is solved as
\bea\label{mu}
&&\mu_0=\sqrt{\frac{2}{1-z_m}}\frac{r_{+}}{l^2}\bigg[3m^2l^2(2-z_m)+\frac{m^4l^4}{2}(1-z_m)\nonumber\\&&-\frac{18+3m^2l^2(1-z_m)}{z_m(\Delta_{+}-2)-\Delta_{+}}\Delta_{+}\bigg]^{1/2}
\bigg\{1-\frac{2\kappa^2}{l^2(1-z_m)}\bigg[\frac{m^2l^2 (1-z_m)\Delta_{+}}{4(2z_m+(1-z_m)\Delta_{+})}\nonumber\\&&+\frac{3\Delta_{+}}{2z_m+(1-z_m)\Delta_{+}}-\frac{3m^2l^2z_m}{4}+m^2l^2\bigg]\bigg\}.\label{mu}
\eea
Without the $\kappa^2$ term, the expression for $\mu_0$ can be reduced to the result of the probe limit case. The $\kappa^2$ term in the above equation is positive, which means that $\mu_0$ increase.
By further using the relation $\mu_0=\frac{\rho}{r_{+}}$, we find an expression for $r_{+}$:
\bea
&&r_{+}=\rho^{1/2}l(\frac{1-z_m}{2})^{1/4}\bigg[3m^2l^2(2-z_m)+\frac{m^4l^4}{2}(1-z_m)\nonumber\\&&-\frac{18+3m^2l^2(1-z_m)}{z_m(\Delta_{+}-2)-\Delta_{+}}\Delta_{+}\bigg]^{-1/4}
\bigg\{1+\frac{2\kappa^2}{l^2(1-z_m)}\bigg[\frac{m^2l^2}{8(2z_m+(1-z_m)\Delta_{+})}\nonumber\\&&+\frac{3\Delta_{+}}{2(2z_m+(1-z_m)\Delta_{+})}-\bigg(\frac{3z_m}{8}-\frac{1}{2}\bigg)m^2l^2\bigg]\bigg\}.\label{rp}
\eea
The Hawking temperature is given by
\be
T=\frac{r_{+}}{4\pi l^2}\bigg(3-\frac{\kappa^2l^2\mu^2_0}{2r^2_{+}}\bigg).\label{t}
\ee
When $\mu_0=\mu_c$, the above Hawking temperature reaches the critical point $T_c$ where the order parameter approaches zero.  From (\ref{mu}) and (\ref{t}), we obtain the critical temperature
\be
T_c=\frac{r_{+}}{4\pi l^2}\bigg\{3-\frac{\kappa^2}{l^2(1-z_m)}\bigg[3m^2l^2(2-z_m)+\frac{m^4l^4}{2}(1-z_m)-\frac{18+3m^2l^2(1-z_m)}{z_m(\Delta_{+}-2)-\Delta_{+}}\Delta_{+}\bigg]\bigg\}
\ee
Together with (\ref{rp}), we write the critical temperature in a form as
\be
T_c=T_1(1-\frac{2\kappa^2}{l^2}T_2)\label{Tc}
\ee
where
\bea
&& T_1=\frac{3\rho^{1/2}}{4\pi l}\bigg(\frac{1-z_m}{2}\bigg)^{\frac{1}{4}}\bigg[3m^2l^2(z_m-2)+\frac{m^4l^4}{2}(1-z_m)
\nonumber\\&&-\frac{18+3m^2l^2(1-z_m)}{z_m(\Delta_{+}-2)-\Delta_{+}}\Delta_{+}\bigg]^{-\frac{1}{4}},\\
&& T_2=\frac{36\Delta_{+}+m^2l^2}{8(1-z_m)[2z_m+(1-z_m)\Delta_{+}]}+\frac{m^2l^2\Delta_{+}}{2[2z_m+(1-z_m)\Delta_{+}]}\nonumber\\&&+\frac{(12-7z_m)m^2l^2}{8(1-z_m)}+\frac{m^4l^4}{12}.
\eea
It is easy to check that when $m^2l^2=-2$, $z_m=1/2$ and thus $\Delta_{+}=2$, we have $T_1=\frac{3 \sqrt{\rho }}{4 \pi l\sqrt{2\sqrt{7}}  }$, the exact result obtained in \cite{gre} for $(2+1)$-dimensional superconductors and $T_2=\frac{5}{6}$. This result is also in good agreement with the numerical result by choosing a proper matching point $z_m$ \cite{horowitz}.
We also find that the corrections due to the backreaction $T_2$ is positive for arbitrary $z_m$ in the region $(0<z_m<1)$.  Therefore, we confirm the result found in \cite{kanno,hartman,barc} that the backreaction makes condensation harder. The reason for the decreasing of the critical temperature can be understood from the relation $T_c \propto 1/{\mu^{1/2}_0}$\cite{gre}. The value of $\mu_0$  increases due to the gravitational backreaction and thus $T_c$ decreases.

\subsection{The upper critical magnetic field with backreaction}
In this section, we will explore the effects of the backreaction on the external critical magnetic field.
{In the neighborhood of the upper critical magnetic field $B_{c2}$, the scalar field
 $\psi$ is small and can be regarded as a perturbation.
The scalar field $\psi$ becomes a function of the bulk coordinate
$z$ and the boundary coordinates $(x,y)$ simultaneously because of the presence of the magnetic field. According
to the AdS/CFT correspondence, if the scalar field $\psi\sim
X(x,y)R(z)$, the vacuum expectation values $<\mathcal {O}> \propto
X(x,y)R(z)$ at the asymptotic AdS boundary (i.e. $z\rightarrow
0$)\cite{aj,ns}. We can simply write $<\mathcal {O}> \propto R(z)$
by dropping the overall factor $X(x,y)$. So, to the leading order, it is
consistent to set the ansatz}
\begin{equation}\label{ft}
A_t=\phi_0(z),~~~A_x=0,~~~A_y=B_{c2}x.
\end{equation}
Note that the applied external magnetic is constant and homogenous. Considering the fact that an external magnetic field is included, one may wonder whether such a constant external magnetic field could backreact on the bulk gravity or not. We may need to consider
the effects of the spatial component of the gauge field in the superconducting phase and assume the gauge field behaves as
\be
A=\phi_0(z)dt+b(z) dx.
\ee
In other words, we need solve the bulk gravity equation for $b(z)$ and the resulted metric is anisotropic. following this line, we may obtain a kind of dyonic black hole solution, which includes charge and magnetism.
However, we notice that several authors have already discussed such conditions in \cite{vector}. In these works, $b(z)$  is interpreted as the vector hair of the black hole. At the AdS boundary, $b(z)$ behaves as
\be
b(z)=\sigma-\xi z+...
\ee
 According the AdS/CFT correspondence, $\xi$ is the dual current density and $\sigma$ is the dual current source of the holographic superfluid. Of course, it is not proper to regard $\xi$ as a homogenous applied magnetic field.
 Actually, when we discuss the vortex structure of the holographic superconductors, in general we should consider
 \be
 \psi_1=\psi_1(x,y,z),~~~A_t=\phi (x,y,z),~~~A_x=A_x(x,y,z),~~~A_y=A_y(x,y,z).
 \ee
 or simply in the polar coordinate $\psi_1=\psi_1(\varrho,z), A_t=\phi_0(\varrho,z), A_{\varphi}=A_{\varphi}(\varrho,z)$ as well as the boundary condition that $A_t(z=1)=0$ and $A_{\varphi}(z=1)$ regular.

 In the presence of external magnetic field, not only the matter fields but also the spacetime metric should  depend on the coordinates $(z,x,y)$. The background static metric may have the form
 \be
 ds^2=g_{00}(z,x,y)dt^2+g_{zz}(z,x,y)dz^2+g_{xx}(z,x,y)dx^2+g_{yy}(z,x,y)dy^2.
 \ee
 In this case, we need solve the Einstein equations
 \be\label{einstein}
 R_{\mu\nu}-\frac{1}{2}g_{\mu\nu}R-\frac{3}{l^2}g_{\mu\nu}=\kappa^2 T_{\mu\nu},
 \ee
 where \be
 T_{\mu\nu}=F_{\mu\lambda}F^{\lambda}_{\nu}-\frac{1}{4}g_{\mu\nu}F^{\lambda \rho}F_{\lambda \rho}
 -g_{\mu\nu}(|D \psi|^2+m^2|\psi|^2)+\bigg[D_{\mu} \psi (D_{\nu} \psi)^{*}+D_{\nu} \psi (D_{\mu} \psi)^{*}\bigg],
 \ee
 together with the Klein-Gordon equation
 \be\label{klein}
\frac{1 }{\sqrt{-g}}D_{\mu}\bigg(\sqrt{-g}g^{\mu\nu}D_{\nu}\psi\bigg)=m^2\psi,
 \ee
 and the Maxwell equation
 \be\label{maxwell}
 \frac{1 }{\sqrt{-g}}\partial_{\lambda}\bigg(\sqrt{-g}g^{\lambda \mu}g^{\rho\sigma}F_{\mu\sigma}\bigg)=g^{\rho\sigma}J_{\sigma},
 \ee
 where we have defined $D_{\mu}\psi=\partial_{\mu}\psi-iA_{\mu}\psi$ and $J_{\sigma}=i[\psi^{*}D_{\sigma}\psi-\psi(D_{\sigma}\psi)^{*}]$.
In this case,  we have three coupled nonlinear partial differential equations involving the metric components, scalar field
$\psi$, the scalar potential $A_t$ and vector potential $\bf{A}$ in which analytic study becomes very difficult to do. Note that we can expand the background geometry in series of $\epsilon$
\bea
g_{\mu\nu}=g_{\mu\nu}^{(0)}+\epsilon^2 g_{\mu\nu}^{(2)}+\epsilon^4 g_{\mu\nu}^{(4)}+...
\eea
To solve these equations analytically we will follow the
logic as shown in table \ref{table}. In the absence of the external magnetic field, the backreaction of the electric field to the background geometry leads to the RNAdS black hole solution at the zeroth order. At the linear order, the metric receives no corrections from matter fields and we need only solve the equation of motion for $\psi_1$ at this moment. As we have done from (\ref{p1}) to (\ref{Tc}), all the equations depends only on the radial coordinate $z$. We obtained the critical temperature with backreaction. When we turn on the external magnetic field, the background spacetime changes because of the presence of $B_{c2}$. We can still expand $\psi$, $A_t$ and $A_{\varphi}$ in series of $\epsilon$. At the leading order, the matter field $\phi_0$ and $A^{(0)}_{\varphi}$ result in a dyonic black hole solution in AdS space. By solving $\psi_1(x_i,z)$ ($x_i=x,y$) at next to leading order, we should obtain the expression for the upper critical magnetic field. The above arguments are actually the logic of the calculation of the whole paper.

\begin{table}[htbp]\label{table}
\begin{center}
\begin{tabular}{|c|c|c|c|c|c|c|c|c|}
\hline
$$&$\rm Vanishing~~~ magnetic~~~ field$&$\rm External~~~ magnetic~~~ field$\\
\hline
$\psi=$&$ \epsilon^1 \psi_1(z)+\epsilon^3 \psi_3+... $&$  \epsilon^1 \psi_1(z,x_i)+\epsilon^3 \psi_3(z,x_i)+...$\\
\hline
$A_t=$&$  \epsilon^0 \phi_0(z)+\epsilon^2 \phi_2(z)+... $&$\epsilon^0 \phi_0(z)+\epsilon^2 \phi_2(z,x_i)+... $\\
\hline
$A_{\varphi}=$&$   0$&$  \epsilon^0 A^{(0)}_{\varphi}(z,x_i)+\epsilon^2 A^{(2)}_{\varphi}(z,x_i)+.. $\\
\hline
$g_{\mu\nu}=$&$  \epsilon^0 g_{RNAdS}+\epsilon^2 g_{\mu\nu}^{(2)} +...$&$  \epsilon^0 g_{dyonic}+\epsilon^2 g_{\mu\nu}^{(2)} +... $\\

\hline
\end{tabular}
\caption{Logic of the analytic calculation. }\label{table}
\end{center}
\end{table}

After justify the usage of  (\ref{ft}), we can then solve the equations of motion order by order. {The black hole carries both electric and magnetic charge  and the bulk Maxwell field yields}
 \be
 A=B_{c2}x dy+\phi  dt
 \ee
At the zeroth order $\mathcal{O}(\epsilon^0)$, we solve the Einstein equation and the line elements of the dyonic black hole metric are given by \cite{dyonic}
\bea
&&ds^2=-f_0dt^2+\frac{r^2_{+}}{z^2l^2}(dx^2+dy^2)+\frac{r^2_{+}}{z^4f_0}dz^2,\\
&&\phi_0=\mu_0-\frac{\rho}{r_{+}}z,~~~ A^{(0)}_y=B_{c2}x
\eea
where $f_0=\frac{r^2_{+}}{z^2l^2}\bigg(1-z\bigg)\bigg(1+z+z^2-\frac{\kappa^2l^2\mu^2_0}{2r^2_{+}}z^3-\frac{\kappa^2l^4B^2_{c2}}{2r^4_{+}}z^3\bigg)$ . The Hawking temperature at the event horizon is evaluated as
\be
T=\frac{r_{+}}{4\pi l^2}\bigg(3-\frac{\kappa^2l^2\mu^2_0}{2r^2_{+}}-\frac{\kappa^2l^2 B^2_{c2}}{2r^4_{+}}\bigg).
\ee

To the linear order, the equation of motion for $\psi_1$ has its new form
\be\label{p2}
f_0\psi''_1+f'_0\psi'_{1}+\frac{r^2_{+}}{z^4}\bigg(\frac{\phi^2_0 }{f_0}-m^2\bigg)\psi_1=-\frac{l^2}{z^2}\bigg[\partial^2_{x}+(\partial_y-iB_{c2}x)^2\bigg]\psi_1,
\ee
where the prime denotes derivative with respect to $z$.
We use separation of variables
\be
\psi_1=e^{ik_y y}X_n(x)R_n(z),
\ee
and obtain the equation of  a two dimensional harmonic oscillator and a equation for $R(z)$
\bea
&&-X''_n(x)+(k_y-B_{c2}x)^2X_n(x)=\lambda_n B_{c2}X_n(x),\label{x}\\
&&f_0R''_n+f'_0R'_{n}+\frac{r^2_{+}}{z^4}\bigg(\frac{\phi^2_0 }{f_0}-m^2\bigg)R_n=\frac{\lambda_n B_{c2}l^2}{z^2}R_n,\label{R},
\eea
where $\lambda_n=2n+1$  is the eigenvalue of the harmonic oscillator equation, $n=0,1,2,...$ denotes the Landau energy level and the prime in (\ref{x}) and (\ref{R}) denote derivative with respect to $x$ and $z$, respectively. The equation (\ref{x})is solved by the Hermite polynomials
\be
X_n(x)=e^{-\frac{(B_{c2}x-k_y)^2}{2B_{c2}}}H_n(x).
\ee  {Let us choose the lowest mode $n=0$ in what follows, which is the first to condensate and is the
most stable solution after condensation}. Actually, the Arikosov vortex lattice is given by a superposition of the lowest energy solutions
\be
\psi_1=R_0(z)\sum_{j} c_j e^{ik_j y}X_0(x),
\ee
where $c_j$ are coefficients that determine the structure of the vortex lattice.

Now we are going to solve  (\ref{R}) by exploring the matching method and find the correction to the upper critical magnetic field away from the probe limit. Again regularity at the horizon requires
\be
R'_0(1)=\frac{m^2r^2_{+}}{f'_0(1)}R_0(1)+\frac{B_{c2}l^2}{f'_0(1)}R_0(1).
\ee
The behavior of $R_0$ at the asymptotic AdS boundary is given by
\be
R_0(z)=C_{+}z^{\Delta_{+}}. \label{R0}
\ee
The scalar potential $\phi_0$
satisfies the boundary condition at the asymptotic AdS region
$\phi_0(z)=\mu-\frac{\rho}{r_+}z$ and vanishes at the horizon
$\phi_0=0,$ as $z\rightarrow 1$. In the strong field limit, the scalar
field $\psi$ is almost vanishing and we can drop out the
$|\psi|^2$ term in the right hand side of equation (\ref{at}). One may find that $\phi_0(z)=\frac{\rho}{r_+} (1-z)$ is a solution
that satisfies (\ref{at}) and the corresponding boundary conditions
\cite{ns}.

In the presence of the external magnetic field, the Taylor expansion of $R_0$ near the horizon still goes as
\be
R_0(z)=R_0(1)-R'_0(1)(1-z)+\frac{1}{2}R''_0(1)(1-z)^2+...\label{Rex}
\ee
From (\ref{R}), we know that near $z=1$, $R''(1)$ is expressed as
\be
R''_0(1)=-\frac{1}{2}\bigg(4+\frac{f''_0(1)}{f'_0(1)}-\frac{m^2r^2_{+}}{f'_0(1)}+\frac{B_{c2}l^2}{f'_0(1)}\bigg)R'_{0}(1)+\frac{B_{c2}l^2}{f'_0(1)}R_0(1)-\frac{r^2_{+}\phi'_0(1)^2}{2 f'^2_0(1)}R_0(1).
\ee
Putting the expressions for $R'_0(1)$ and $R''_0(1)$ into (\ref{Rex}), we obtain
\bea
R_0(z)&=&R_0(1)-\bigg(\frac{r^2_{+}m^2}{f'_0(1)}+\frac{B_{c2}l^2}{f'_0(1)}\bigg)R_0(1)(1-z)+\bigg[-\frac{r^2_{+}m^2+B_{c2}l^2}{4f'_0(1)}\bigg(4+\frac{f''_0(1)}{f'_0(1)}\nonumber\\&-&\frac{r^2_{+}m^2+B_{c2}l^2}{f'_0(1)}\bigg)
+\frac{B_{c2}l^2}{2f'_0(1)}-\frac{r^2_{+}}{4}\frac{\phi'_1(1)^2}{f'_0(1)^2}\bigg]R_0(1)(1-z)^2+...\label{R01}
\eea
We  connect the two solutions (\ref{R0}) and (\ref{R01}) at a intermediate point $z_m$ smoothly and thus find that
\bea
C_{+}z^{\Delta_{+}}_m&=&R_0(1)-\bigg(\frac{r^2_{+}m^2}{f'_0(1)}+\frac{B_{c2}l^2}{f'_0(1)}\bigg)R_0(1)(1-z_m)+\bigg[-\frac{r^2_{+}m^2+B_{c2}l^2}{4f'_0(1)}\bigg(4+\frac{f''_0(1)}{f'_0(1)}\nonumber\\&-&\frac{r^2_{+}m^2+B_{c2}l^2}{f'_0(1)}\bigg)
+\frac{B_{c2}l^2}{2f'_0(1)}-\frac{r^2_{+}}{4}\frac{\phi'_1(1)^2}{f'_0(1)^2}\bigg]R_0(1)(1-z_m)^2,\\
\Delta_{+}z^{\Delta_{+}-1}_mC_{+}&=&\frac{r^2_{+}m^2+B_{c2}l^2}{f'_0(1)}R_0(1)-2\bigg[-\frac{r^2_{+}m^2+B_{c2}l^2}{4f'_0(1)}\bigg(4+\frac{f''_0(1)}{f'_0(1)}-\frac{r^2_{+}m^2+B_{c2}l^2}{f'_0(1)}\bigg)
\nonumber\\&&+\frac{B_{c2}l^2}{2f'_0(1)}-\frac{r^2_{+}}{4}\frac{\phi'_1(1)^2}{f'_0(1)^2}\bigg]R_0(1)(1-z_m).\label{R2}
\eea
From the above equations, we find that
\be
C_{+}=\frac{2z_m}{2z_m+(1-z_m)\Delta_{+}}z^{-\Delta_{+}}_m\bigg(1-\frac{1-z_m}{2}\frac{r^2_{+}m^2+B_{c2}l^2}{f'_0(1)}\bigg)R_0(1)
\ee
Substituting the above relation back into  (\ref{R2}), we get a non-trivial expression
\bea\label{sf}
&&\frac{2\Delta_{+}}{2z_m+(1-z_m)\Delta_{+}}-
\bigg[\frac{(1-z_m)\Delta_{+}}{2z_m+(1-z_m)\Delta_{+}}+(3-2z_m)\bigg]\frac{r^2_{+}m^2+B_{c2}l^2}{f'_0(1)}\nonumber\\&-&\frac{(1-z_m)(r^2_{+}m^2+B_{c2}l^2)}{2}\frac{f''_0(1)}{f'_0(1)^2}
+\frac{1-z_m}{2}\frac{(r^2_{+}m^2+B_{c2}l^2)^2}{f'_0(1)^2}\nonumber\\&+&\frac{B_{c2}l^2}{f'_0(1)}(1-z_m)-\frac{(1-z_m)r^2_{+}}{2}\frac{\phi'_0(1)^2}{f'_0(1)^2}=0.\label{re}
\eea
When we turn off the magnetic field $B_{c2}=0$, (\ref{re}) returns to (\ref{re1}). In the presence of the magnetic field, both the charge and the magnetic field can backreact on the black hole. The critical temperature should receive further corrections  from the magnetic field.  The difference between (\ref{re}) and (\ref{main}) comes from the $B_{c2}$ related terms, which goes as
\bea
&&\frac{2\Delta_{+}}{2z_m+(1-z_m)}3\kappa^2B^2_{c2}+\bigg(3-2z_m+\frac{(z_m-1)\Delta_{+}}{z_m(\Delta_{+}-2)-\Delta_{+}}\bigg)\bigg(\frac{m^2l^2B^2_{c2}}{2}\kappa^2+\frac{B^3_{c2}\kappa^2}{2r^2_{+}}\nonumber\\&-&3B_{c2}r^2_{+}\bigg)
+\frac{1-z_m}{2}\bigg(\frac{m^2l^2B^2_{c2}}\kappa^2+\frac{B^3_{c2}\kappa^2}{r^2_{+}}+6B_{c2}r^2_{+}-B^2_{c2}+2m^2l^2r^2_{+}B_{c2}\bigg)\nonumber\\&-&B_{c2}(1-z_m)\bigg(\frac{\kappa^2B^2_{c2}}{2r^2_{+}}-3r^2_{+}\bigg)
=\bigg(3-2z_m+\frac{(z_m-1)\Delta_{+}}{z_m(\Delta_{+}-2)-\Delta_{+}}\bigg)\frac{B_{c2}\kappa^2}{2}\mu^2_0.
\eea
We obtain a relation between $B_{c2}$ and $r_{+}$ by using (\ref{mu}) and $m^2l^2=-2$, $z_m=1/2$, $q=1$,  $\Delta_{+}=2$,
\bea\label{bbc}
B_{c2}=\frac{58}{5}r^2_{+}-812 \kappa ^2r^2_{+}+\mathcal{O}(\kappa^4).
\eea
The critical temperature dropped because of the magnetic field
\be\label{ntc}
T_c=T_1(1-\frac{1807}{75}\frac{\kappa^2}{l^2}),
\ee
where $T_1=\frac{3 \sqrt{\rho }}{4 \pi l\sqrt{2\sqrt{7}}  }$. This reflects the fact that condensation becomes even harder when one turns on the external magnetic field.

Note that (\ref{bbc}) is not enough to determine the relation among the upper critical magnetic field, the system temperature $T$ and  the critical temperature $T_c$.
Considering the values of $f'_0(1)$, $f''_0(1)$ and $\phi'_0(1)$ and solving  (\ref{sf}) to the first order of $\kappa^2$, we get
\bea
&&\mu_0=\frac{\mathcal{H}}{\sqrt{1-z_m}l^2}\bigg\{1+\frac{\kappa^2}{2l^2}\bigg(\frac{1}{1-z_m}-\frac{B^2_{c2}l^6}{\mathcal{H}r^4_{+}}\bigg)
\bigg(12\Delta_{+}+2B_{c2}l^4\bigg(z_m(3+z_m(\Delta_{+}-2)-3\Delta_{+})\nonumber\\&&+2\Delta_{+}+m^2l^2(z(8+3z_m(\Delta_{+}-2)-8\Delta_{+})
+5\Delta_{+})/r^2_{+}\bigg)[z_m(\Delta_{+}-2)-\Delta_{+}]^{-1}\bigg)\bigg\}\nonumber\\&&+\mathcal{O}(\kappa^4),
\eea
with
\bea
\mathcal{H}&=&\bigg\{\frac{B^2_{c2}}{r^4_{+}}l^8(1-z_m)+\frac{36\Delta_{+}}{z_m(\Delta_{+}-2)-\Delta_{+}}+\frac{6m^2l^2\bigg(z_m(4+z_m(\Delta_{+}-2)-4\Delta_{+})+3\Delta_{+}\bigg)}{z_m(\Delta_{+}-2)-\Delta_{+}}
\nonumber\\&+&m^4l^4(z_m-1)+\frac{12B_{c2}(z_m+\Delta_{+}(1-z_m))}{r^2_{+}(z_m(\Delta_{+}-2)-\Delta_{+})}\bigg\}
[z_m(\Delta_{+}-2)-\Delta_{+}]^{-1}.
\eea
Combining the above equation with $\mu_0=\frac{\rho}{r_{+}}$, we can obtain a relation between $B_{c2}$ and $r_{+}$.  We then substitute (\ref{mu}) and (\ref{bbc}) into the Hawking temperature $T=\frac{r_{+}}{4\pi l^2}\bigg(3-\frac{\kappa^2l^2\mu^2_0}{2r^2_{+}}-\frac{\kappa^2l^2 B^2_{c2}}{2r^4_{+}}\bigg)$ and now the Hawking temperature plays the role of the critical temperature in the presence of magnetic fields. In order to have a clear picture,
by choosing $m^2l^2=-2$, $\Delta_{+}=2$ and $z_m=1/2$ and further using the relation between $r_{+}$ and $T$ from the new Hawking temperature,
we find that  the upper critical magnetic field $B_{c2}$ yields
\bea
B_{c2}&=&B_1+B_2 \kappa^2 =\frac{1}{9}\bigg(\sqrt{5376\pi^4T^4+\frac{81\rho^2}{l^4}}-112\pi^2T^2\bigg)\nonumber\\
&+& \kappa ^2\bigg[\frac{14565376 \pi ^4 T^6-46575 T^2 \rho ^2}{450 \sqrt{5376 \pi ^4 T^8+81 T^4 \rho ^2}}-\frac{455168}{675} \pi ^2 T^2
\nonumber\\&+&\frac{45 \rho ^2}{32 \pi ^2 T^2}\bigg]+\mathcal{O}(\kappa^4).
\eea
\begin{figure}[htbp]
 \begin{minipage}{1\hsize}
\begin{center}
%\vspace*{10mm}
\includegraphics*[scale=0.6] {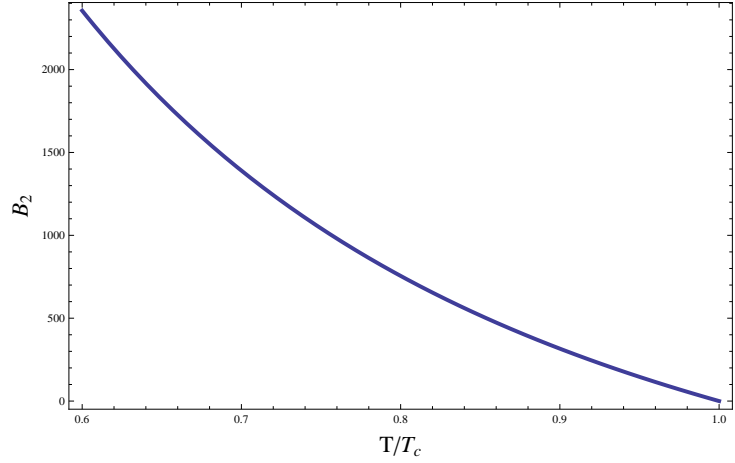}
\end{center}
\caption{(color online) The coefficient of the $\kappa^2$ term of the upper critical magnetic field as a function of the temperature $T/T_c$. We set  $T_c=1$ here} \label{B1}
\end{minipage}
\end{figure}
In this case, the charge density  can be evaluated from (\ref{ntc}), that is $\rho=\frac{32 \sqrt{7} \pi ^2}{9}T^2_c \bigg(1-\frac{2\kappa^2}{l^2}T_2\bigg)^{-2}$.
The upper critical magnetic field $B_{c2}$ in series of $\kappa^2$  can be expressed as
\bea\label{bc}
B_{c2}&=&\frac{16}{9}T^2_c \pi ^2 \bigg[\left(\sqrt{7}\sqrt{4+3 \frac{T^4}{T^4_c}}-7\frac{ T^2}{T^2_c}\right)+\bigg(\frac{\frac{4064}{25}\frac{ T^4}{T^4_c}+\frac{5503}{75}}{\sqrt{4+3 \frac{T^4}{T^4_c}}}\sqrt{7}+70\frac{T^2_{c}}{T^2}\nonumber\\&-&\frac{28448}{75} \frac{T^2}{T^2_{c}}\bigg) \kappa^2+\mathcal{O}(\kappa^4)\bigg].
\eea

\begin{figure}[htbp]
 \begin{minipage}{1\hsize}
\begin{center}
%\vspace*{10mm}
\includegraphics*[scale=0.8] {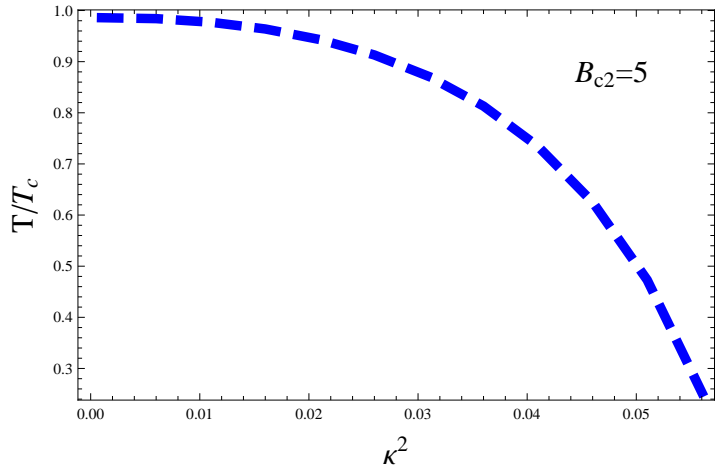}
\end{center}
\caption{(color online) For fixed external magnetic field, the phase transition temperature decreases if $\kappa^2$ becomes larger. We set  $T_c=1$ here.} \label{kt}
\end{minipage}
\end{figure}
It is worth noting that $T_c$ means the critical temperature without magnetic fields and gravitational backreaction. The result (\ref{bc}) also implies that it is only applicable near the critical temperature $T_c$ because the $\kappa^2$ term will be divergent  in the low temperature limit. One may find that when $\kappa^2=0$, the result exactly agrees with \cite{gw}, which is also consistent with the Ginzburg-Landau theory where $B_{c2}\propto \bigg(1-T/T_c\bigg)$. We also find that the coefficient of the $\kappa^2$ term is positive for the system temperature $T$ (see Fig. \ref{B1}).
 This result indicates that the effects of the backreaction enhance the value of the upper critical magnetic field. The increasing of the magnetic field $B_{c2}$ can be explained from the relation that $B_{c2} \propto \mu_0$
 for fixed value of $r_{+}$ \cite{gw}.  Therefore, if the value of $\mu_0$ becomes larger, then $B_{c2}$ increases. However, this does not mean condensation become easy in the presence of the magnetic field. We can see from Fig. \ref{kt} that for fixed magnetic field, the phase transition temperature goes down as $\kappa^2$ increases.

\subsection{Numerical results}

 \begin{table}
\centering
\begin{tabular}{|c|c|c|c|c|c|c|c|c|}
  \hline
  $\kappa^2$ & 0 & 0.025& 0.05 & 0.1 & 0.15 & 0.2 & 0.3 & 0.35 \\ \hline
  ${T}/{\sqrt{\rho}}$ & 0.118 & 0.111& 0.104 & 0.09& 0.07&0.06& 0.03&0.01 \\ \hline
 \end{tabular}
\caption{The critical temperature $T$ drops as $\kappa^2$ increases in the absence of the magnetic field.}
\label{tableO}
\end{table}

\begin{figure}[htbp]
 \begin{minipage}{1\hsize}
\begin{center}
%\vspace*{10mm}
\includegraphics*[scale=0.8] {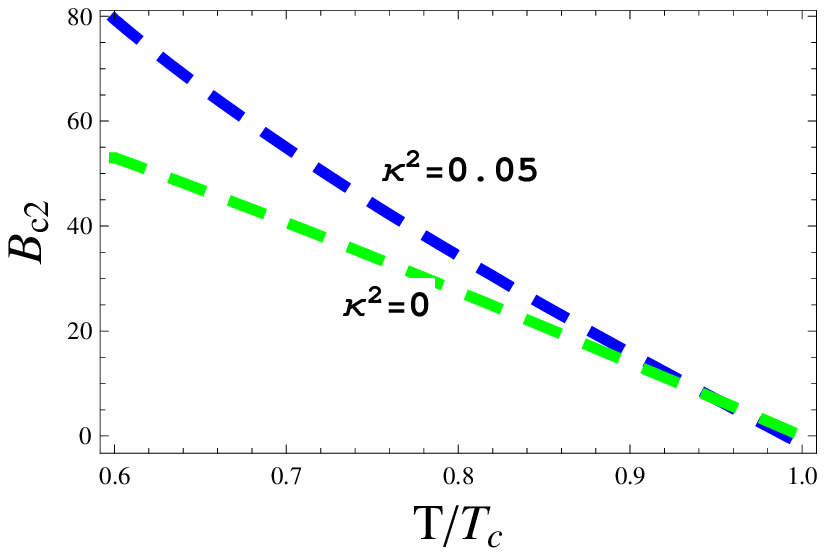}
\includegraphics*[scale=0.8] {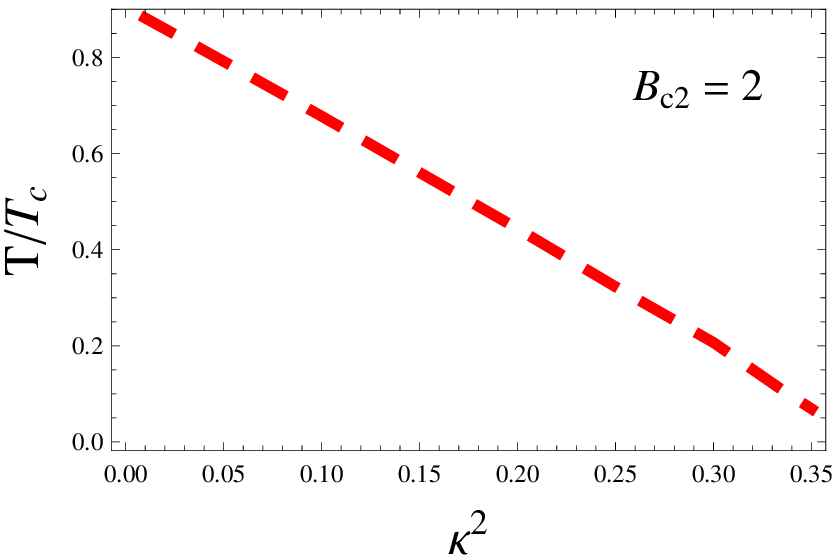}
\end{center}
\caption{(color online) Left: The external  magnetic field as a function of $T/T_c$ at $\kappa^2=0$ (Green) and $\kappa^2=0.01$ (Blue). Right: The phase transition temperature decreases if $\kappa^2$ increases in the case that $B_{c2}=2$. In both cases, we  choose $T_c=1$. } \label{BCT}
\end{minipage}
\end{figure}
For completeness of our study, we carry on numerical computation in this subsection. We first solve the equations of motion (\ref{m1}) to(\ref{p1}) in the absence of the external magnetic field and  from which we can obtain the critical temperature and the phase diagram. The properties of holographic superconductors without magnetic fields away from the probe limit were studied numerically in \cite{wang4} by setting $2\kappa^2=1$ and finite $q$. We work in the case $q=1$ but finite $\kappa^2$ instead and set $r_{+}=1$ and $l=1$ in the numerical computation. The critical temperature $T$ as a function of the backreaction $\kappa^2$ is shown in Table \ref{tableO}. It is clear that  the critical temperature $T$ drops as $\kappa^2$ increases, which is in consistent with \cite{wang4,hartman}.

We then consider the behavior of the external magnetic field numerically to the linear level by solving equation  (\ref{p2}). In Fig.\ref{BCT} (left), we find that the external magnetic field drops  in  different ways for  $\kappa^2=0$  case and $\kappa^2=0.01$ case. This is in consistent with the analytic calculation in the range $T\sim T_c$. That is to say, although the critical temperature is significantly suppressed by a non-zero $\kappa^2$, the upper bound of $B_{c2}$ become larger.
  In Fig.\ref{BCT} (right), we also plot the phase diagram of the critical
 temperature against the gravitational backreaction. When we fix the magnetic field, the phase transition temperature is depressed as $\kappa^2$ increases, which is also comparable with the analytic results at qualitative level since the  analytic method closely depends on the matching point.
 Note that the numerical results presented here can be regarded as a side note because we mainly deal with analytical calculation in this paper. A more general and thorough numerical computation in the presence of external magnetic field with backreaction is called for in the future.

\section{Conclusion}

In this paper, we have investigated the effect of backreaction to the upper critical magnetic field of $(2+1)$-dimensional holographic superconductors in Einstein gravity by using the analytical method developed in \cite{kanno,gre}. As a consistent check, we have derived the critical temperature with backreaction in four-dimensional Einstein gravity and confirmed the numerical result given in \cite{wang4} that backreaction makes the condensation harder to form. We have obtained the spatially dependent condensate solutions in the presence of the magnetism. The coefficient of spacetime backreaction on the upper critical magnetic field is positive for  Einstein gravity, which indicates that the magnetic field becomes strong with respect to the backreaction in consistent with reference \cite{gtw} . We have also shown the corresponding numerical results for each case.

We can see that the spacetime backreaction presents us an interesting property of holographic superconductors: While the backreaction causes the depression of the critical temperature, it can enhance the upper critical magnetic field. The upper critical field  $B_{c2}$ is an important parameter because it determines the value of the coherence length and strongly affects the  critical current density  $J_c$. The improvement in $B_{c2}$
has been the main research topic for some experiments. In this paper, we  work in the small backreaction limit (i.e. $\kappa^2\ll 1$). So if we regard the backreaction as a factor of ``doping'' in holographic superconductors, then we may find that  the backreaction plays the same role as carbon doping in $\rm MgB_{2}$ reported in recent experiments\cite{mgb}: It results in the depression in $T_c$, while the $B_{c2}$ performance is improved. Otherwise, we can treat the probe limit approximation as the ``effective doping'': comparing with the superconducting properties with backreaction, the probe limit approximation improves the critical temperature but reduces the upper critical magnetic field.
  In microscopic models of high temperature superconductors, the interaction between  doping and electrons contributes a potential term in the Hamiltonian  and the self energy of the superconducting quasi-particles will be changed. The self-energy can be
 calculated by using the Green function and the correction to the critical temperature can be read off from the Green function\cite{poole}.  For holographic superconductors, the effective mass term is changed with the variation of $\kappa^2$.
 The extension of this work to the five-dimensional Gauss-Bonnet gravity case would be interesting\cite{g4,gsst,ge,bm,esc,ges}. But since the resulting metric is anisotropic and analytic calculation become very difficult and involving, we would like to leave it to the future publication by using numerical calculation.

\section*{Acknowledgements}
XHG would like to thank Ying Jiang for helpful discussions on $\rm MgB_2$. The work   was partly supported by NSFC, China
(No. 11005072),  Shanghai Rising-Star Program and
Shanghai Leading
Academic Discipline Project (S30105).


\begin{thebibliography}{99}

\bibitem{ads/cft}
J. M. Maldacena, {Adv. Theor. Math. Phys.}  2 (1998) 231, {
[arXiv:hep-th/9711200]}.

\bibitem{gkp}
S. S. Gubser, I.R. Klebanov and A.M. Polyakov, Phys.\ Lett.\ {
B428} (1998) 105, { [arXiv:hep-th/9802109]}.

\bibitem{w}
E. Witten, Adv.\ Theor.\ Math.\ Phys.\ { 2} (1998) 253, {
[arXiv:hep-th/9802150]}.
%\cite{Gubser:2005ih}

\bibitem{gub1}
  S.~S.~Gubser,
  %``Phase transitions near black hole horizons,''
  Class.\ Quant.\ Grav.\  { 22}  (2005) 5121.
  %[arXiv:hep-th/0505189].
    %%CITATION = CQGRD,22,5121;%%

\bibitem{gub2}
S. S. Gubser, Phys. Rev. D { 78} (2008)   065034.

\bibitem{3h} C. P. Herzog, J. Phys. A { 42} (2009) 343001 [arXiv:0904.1975[hep-th]].
 S. A. Hartnoll, Class. Quant. Grav. { 26} (2009) 224002 [arXiv:0903.3246[hep-th]].

\bibitem{horowitz}
S. A. Hartnoll, C. P. Herzog, and G. T. Horowitz, Phys. Rev. Lett.
101 (2008)  031601 [arXiv:0803.3295[hep-th]].

\bibitem{horowitz2} S. A. Hartnoll, C. P. Herzog, and G. T. Horowitz,  J. High Energy Phys. { 0812}   (2008) 015[arXiv:0810.1563[hep-th]].
\bibitem{gre}
R. Gregory, S. Kanno and J. Soda, J. High Energy Phys. { 10}
(2009)  010 [arXiv:0907.3203[hep-th]]
\bibitem{wang1}
Q. Pan, Bin Wang, E. Papantonopoulos, J. Oliveria and A. B. Pavan,%Holographic Superconductors with various condensates in Einstein-Gauss-Bonnet gravity
Phys.Rev.D { 81}, (2010) 106007 [arXiv:0912.2475[hep-th]];

\bibitem{wang2} Q. Pan
and B. Wang, Phys.Lett. { B693}   (2010) 159
 [arXiv:1005.4743 [hep-th]]

 \bibitem{wang3}Y. Liu, Q. Pan, B. Wang and R. G. Cai,  Phys. Lett. { B693}   (2010) 343 [arXiv:1007.2536 [hep-th]].

 \bibitem{wang4} Q. Pan and B. Wang [arXiv:1101.0222 [hep-th]
]

\bibitem{siani} M. Siani,
%Holographic Superconductors and Higher Curvature Corrections
J. High Energy Phys. { 12} (2010) 035
\bibitem{wang5}Y. Peng,  Q. Pan and B. Wang
%Various types of phase transitions in the AdS soliton background
[arXiv:1104.2478[hep-th]]

\bibitem{cai} R. G. Cai, Z. Y. Nie and H. Q. Zhang, Phys.Rev. { D 82}  (2010) 06607
; R. G. Cai, Z. Y. Nie and H. Q. Zhang, Phys.Rev.{ D 83} (2011) 06613


\bibitem{ling} X. M. Kuang, W. J. Li and Y. Ling,
%Holographic Superconductors in Quasi-topological Gravity
J. High Energy Phys. { 1012}  (2010) 069 [arXiv:1008.4066 [hep-th]]

\bibitem{kuang} J. P. Wu, Y. Cao, X. M. Kuang and W. J. Li,
%The 3+1 holographic superconductor with Weyl corrections
Phys. Lett. { B697}  (2011) 153 [arXiv:1010.1929 [hep-th]]


\bibitem{hartman} Y. Brihaye and B. Hartman, Phys. Rev. D { 81} (2010) 126008

\bibitem{barc} L. Barcaly, R. Gregory, S. Kanno and P. Sutcliffe,
% Gauss-Bonnet Holographic Superconductors
J. High Energy Phys. 1012  (2010) 029

\bibitem{kanno} S. Kanno, % A note on Gauss-Bonnet holographic superconductors
 Class.Quant.Grav.28 (2011) 127001 [arXiv:1103.5022[hep-th]].


\bibitem{jing} J. Jing, L. Wang, Q. Pan and S. Chen,
%Holographic Superconductors in Gauss-Bonnet gravity with Born-Infeld electrodynamics
 Phys. Rev. { D 83}  (2011) 066010
[arXiv:1012.0644 [gr-qc]]; S. Chen, Q. Pan and J. Jing,
%Holographic superconductor models in the non-minimal derivative coupling theory
[ arXiv:1012.3820[hep-th]].
\bibitem{chen} C. M. Chen and W. F. Wu,
%An Analytic Analysis of Phase Transitions in Holographic Superconductors
[arXiv:1103.5130[hep-th]].

\bibitem{amm1}
M. Ammon, J. Erdmenger, V. Grass, P. Kerner and A. O’Bannon, %On Holographic p-wave Superfluids with Back-reaction
Phys. Lett. B { 686} (2010)
192 [arXiv:0912.3515 [hep-th]].
\bibitem{aj}
T. Albash and C. V. Johnson,
%“A Holographic Superconductor in an External Magnetic Field,”
J. High Energy Phys. { 0809}  (2008) 121 [arXiv:0804.3466 [hep-th]]


\bibitem{wk} E. Nakano, W.Y. Wen, Phys. Rev. D 78 (2008) 046004.

\bibitem{amm}
 A. Ammon, J.
Erdmenger, M. Kaminski and P. Kerner,
%Flavor Superconductivity from Gauge/Gravity Duality
 J. High Energy Phys. 0910 (2009)  067

\bibitem{john}
 T. Albash and C. V. Johnson, %“Vortex and Droplet Engineering in Holographic Superconductors,”
  arXiv:0906.1795
[hep-th].

\bibitem{mon}
M. Montull, A. Pomarol, and P. J. Silva, Phys. Rev. Lett { 103}
 (2009) 091601.

\bibitem{ns} K. Maeda, M. Natsuume and T. Okamura, %“ Vortex lattice for a holographic superconductor,”
  Phys. Rev. { D 81}  (2010) 026002.

  \bibitem{silva} O. Domenech, M. Montull, A. Pomarol and A. Salvio and P. J. Silva,
  %Emergent Gauge Fields in Holographic Superconductors
    J. High Energy Phys. { 1008}  (2010) 033.

    \bibitem{gw}
X. H. Ge, B. Wang, S. F. Wu and G. H. Yang,
%Analytical study on holographic superconductors in external magnetic field
J. High Energy Phys. { 1008} (2010) 108 [arXiv:1002.4901 [hep-th]]

\bibitem{wu} J. P. Wu, %The Stuckelberg Holographic Superconductors in Constant External Magnetic Field,
  [arXiv:1006.0456 [hep-th]]

\bibitem{queen} G. Tallarita and  S. Thomas,
%Maxwell-Chern-Simons Vortices and Holographic Superconductors.
J. High Energy Phys. { 1012}   (2010) 090

\bibitem{mann}R. B. Mann, R. Pourhasan
%Gauss-Bonnet Black Holes and Heavy Fermion Metals
[arXiv:1105.0389 [hep-th]].
\bibitem{vector} D. Arean, P. Basu and C. Krishnan,
JHEP 10 (2010) 006
\bibitem{BF} P. Breitenloher and D. Z. Freedman, Ann. Phys. { 144} 249 (1982).
\bibitem{dyonic} L. J. Romans, Nucl. Phys. B { 383} 395 (1992) [arXiv:hep-th/9203018]
\bibitem{gtw}X. H. Ge, S. F. Tu and B. Wang
%d-wave holographic supercondcutors with backreaction in external magnetic field
J. High Energy Phys. { 1209} (2012) 088 [arXiv:1209.4272].
\bibitem{mgb} Y. M. Ma, X. P. Zhang, G. Nishijima, K. Watanabe, S. Awaji and X. D. Bai, %Significantly enhanced critical current densities in MgB2 tapes made by a scaleable nanocarbon addition route
App. Phys. Lett. { 88} (2006) 072502; Y. Zhang, S. H. Zhou, C. Lu, K. Konstantinov and S. X. Dou, %The effect of carbon doping on the upper critical ?eld (Hc2) and resistivity of MgB2 by using sucrose (C12H22O11)asthe carbon source
Supercond. Sci. Technol. { 22} (2009) 015025; Xianping Zhang et al  %Doping with a special carbohydrate, C9H11NO, to improve the Jc–B properties of MgB2 tapes
 Supercond. Sci. Technol. { 23} (2010) 025024.
 \bibitem{poole} C. P. Poole, H. A. Farach and R. J. Creswick, ``
Superconductivity'', Academic Press, Netherlands,  (2007) .
\bibitem{g4}
I. P. Neupane, Phys. Rev. { D67} (2003) 061501,
[arXiv:hep-th/0212092].
\bibitem{gsst} X.~H. Ge, Y. Matsuo, F.-W. Shu, S.-J. Sin and T. Tsukioka, J. High Energy Phys. {
10}, 009 (2008)  { [arXiv:0808.2354[hep-th]]}
\bibitem{ge}
X.~H. Ge and S.-J. Sin, J. High Energy Phys. { 05}, 051 (2009) {
[arXiv:0903.2527[hep-th]]}
\bibitem{bm} A. Buchel and R. C. Myers,
% Causality of holographic hydrodynamics
J. High Energy Phys. { 08}, 016 (2009) {
[arXiv:0906.2922[hep-th]]}
\bibitem{esc} A. Buchel, J. Escobedo, R. C. Myers, M. F. Paulos, A. Sinha, M.
Smolkin, %Holographic GB gravity in arbitrary dimensions
J. High Energy Phys. { 03} 111 (2010) { [arXiv:0911.4257
[hep-th]]}
\bibitem{ges}
X.~H. Ge, S. J. Sin, S. F. Wu and G. H. Yang, Phys. Rev. D { 80},
104019 (2009) { [arXiv:0905.2675[hep-th]]}

\end{thebibliography}
\end{document}